\documentclass[final]{svjour3}
\usepackage{graphicx}
\usepackage{rotating}
\usepackage{amssymb}
\usepackage{mathptmx}
\usepackage{color}
\usepackage{upgreek}
\usepackage[numbers]{natbib}
\usepackage[colorlinks=true,allcolors=blue]{hyperref}
\newcommand{\doi}[1]{{\href{https://doi.org/#1}{DOI:#1}}}

\makeatletter
\journalname{Journal of Low Temperature Physics}

\newcommand{\com}[1]{}

\begin{document}

\newcommand{\hdblarrow}{H\makebox[0.9ex][l]{$\downdownarrows$}-}
\title{First Operation of TES Microcalorimeters in Space with the Micro-X Sounding Rocket}

\author{J.S. Adams \and R. Baker \and S.R. Bandler \and N. Bastidon$^*$ \and M.E. Danowski \and W.B. Doriese \and M.E. Eckart \and E. Figueroa-Feliciano \and D.C. Goldfinger$^*$ \and   S.N.T. Heine \and G.C. Hilton \and A.J.F. Hubbard \and R.L. Kelley \and C.A. Kilbourne \and R.E. Manzagol-Harwood \and D. McCammon \and T. Okajima \and F.S. Porter \and C.D. Reintsema \and P. Serlemitsos \and S.J. Smith \and J.N. Ullom \and P. Wikus}

\institute{$^*$Corresponding authors:\\
\email{bastidon.microx@gmail.com}\\
\email{dgoldfin@mit.edu}\\
Department of Physics and Astronomy, Northwestern University,\\ Evanston, IL 60208, United States
}

\maketitle

\begin{abstract}

Micro-X is a sounding rocket-borne instrument that uses a microcalorimeter array to perform high-resolution X-ray spectroscopy. This instrument flew for the first time on July 22\textsuperscript{nd}, 2018 from the White Sands Missile Range, USA. This flight marks the first successful operation of a Transition-Edge Sensor array and its time division multiplexing read-out system in space. This launch was dedicated to the observation of the supernova remnant Cassiopeia A. A failure in the attitude control system prevented the rocket from pointing and led to no time on target.  The on-board calibration source provided X-rays in flight, and it is used to compare detector performance during pre-flight integration, flight, and after the successful post-flight recovery. This calibration data demonstrates the capabilities of the detector in a space environment as well as its potential for future flights.


\end{abstract}

\section{Introduction}
\begin{figure}
	\begin{center}
		\includegraphics[width=0.9\linewidth,keepaspectratio]{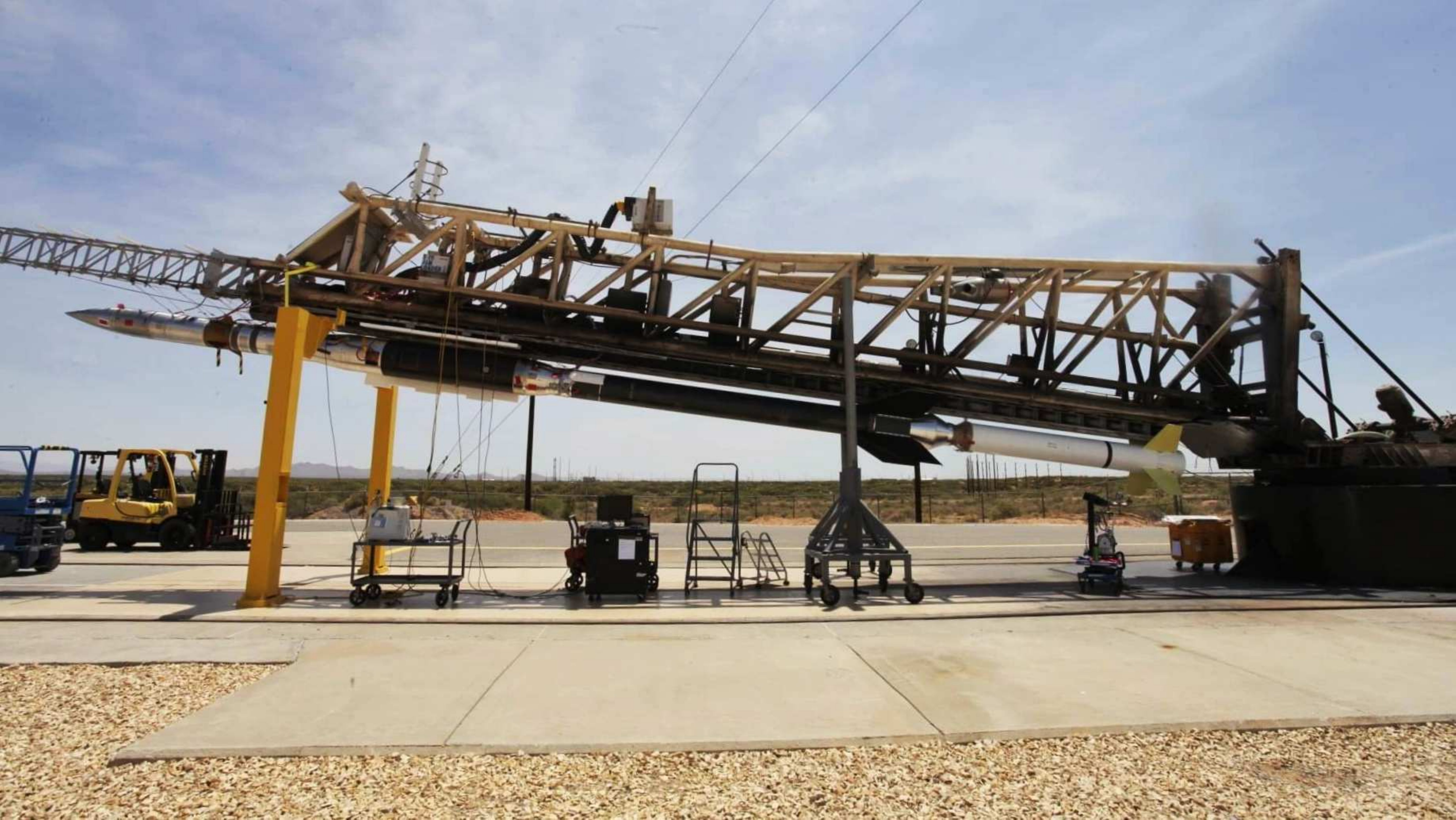}
		\caption{\label{MicroXrail} Micro-X rocket in horizontal position on the rail at the White Sands Missile Range, July 2018. (Color figure online.)}
	\end{center}
\end{figure}

The Micro-X sounding rocket X-ray imaging spectrometer (Fig. \ref{MicroXrail}) was launched for the first time on the night of July 22\textsuperscript{nd}, 2018 from the White Sands Missile Range (New Mexico, USA). The aim was to demonstrate the use of Transition Edge Sensor (TES) microcalorimeters with Superconducting Quantum Interference Device (SQUID) readout using time-division multiplexing in space \cite{Wikus2010, Irwin2005}.  The science goal was to obtain a high resolution spectrum of the Cassiopeia A supernova remnant\footnote{The supernova remnant considered for this mission depends on the time of the year. The primary target (Puppis A) is not visible during the summer, so the secondary target was chosen for the first flight.}. Micro-X uses TES devices developed at NASA GSFC with intrinsic energy resolution of 4.5~eV over the energy range of interest (0.1~-~2.5~keV) \cite{Goldfinger2016}.
	
The Micro-X detector array is comprised of 128 individual TES detector/pixels, each with a 590~$\upmu$m x 590~$\upmu$m absorber organized in a 12 by 12 square grid at a 600~$\upmu$m pitch. The readout system determines the number of channels, so we instrument a circular pattern with an 11.8~arcmin total field of view, omitting the corner pixels \cite{Wikus2010}. The sensors and their support stems are suspended on a 1~$\upmu$m silicon nitride membrane \cite{Eckart2009}. The X-ray absorbers have 3.4 $\upmu$m of bismuth on top of 0.6 $\upmu$m of gold, designed to achieve high spectral resolution and near-unity quantum efficiency over the science bandpass while remaining mechanically robust. The molybdenum/gold bilayer TESs have a transition temperature of T$_{c}$=120~mK. 

The detectors are cooled by an Adiabatic Demagnetization Refrigerator (ADR) to 75~mK with a hold time of 9 hours \cite{Goldfinger2016thermal}. A system of three vibration isolation stages with progressively higher resonance frequencies decouple the ADR salt pill and detector stage from launch vibrations \cite{Danowski2016}. These are comprised of mechanical isolation of the cryostat from the rocket skin by six wire rope isolators divided in two sets (top and bottom); decoupling of the helium tank inside the cryostat from the outer cryostat skin by a spring mount fixture; and isolation of the detector stage from the helium tank via kevlar suspension.  

The detectors are read out with time-division multiplexing using a three-stage SQUID amplifier (designed by NIST in 2006) with 8 columns of 16 rows each \cite{Korte2003, Stiehl2010}. The readout consists of two independent sets of 4 columns each with a sampling frequency of 65~kHz per pixel, decimated to 21.7~kHz. To reduce the ambient magnetic field seen by the detector system, the SQUIDs and the TESs are placed in a superconducting niobium enclosure with openings for the X-ray photons, the thermal bus and readout wiring. To prevent trapped flux, a Metglas\textsuperscript{\textregistered}\com{not for today, but let's add a citation to the paper we used to make the Metglas Blanket} shield is wrapped around the rocket skin before the initial cooldown and removed prior to launch. The ADR magnet is equipped with a bucking coil to reduce its field near the detectors.

A two-stage conically approximated Wolter I mirror, previously flown on the SXS rocket payload, with a focal length of 2.1~m is used to focus the image on the detector plane \cite{Wikus2010,Aschenbach2009,Petre2010}. The mirror provides an effective area of 300~cm$^2$ at 1~keV, and a 2.4" point spread function.  In flight, the rocket pointing is maintained using a celestial Attitude Control System (ACS) constituted of a ST-5000 star tracker \cite{Percival2008} with bi-level cold gas thrusters.  Four mesh optical/infrared filters (28~mm in diameter, 250~$\AA$ Al, 500~$\AA$ polyimide) and two non-mesh filters (15~mm in diameter, 200~$\AA$ Al, 500~$\AA$ polyimide) are placed across the X-ray aperture to reduce radiative loading\footnote{The silicon mesh provides mechanical support to the aluminum and polyimide films for the filters with larger diameters.} \cite{SarahThesis}.  Inside the cryostat, a potassium chloride crystal illuminated by the 5.89~keV \textsuperscript{55}Mn K-$\alpha$ line from an \textsuperscript{55}Fe source is used to calibrate the detectors, both in flight and on the ground \cite{SarahThesis, DavidThesis, Lopez2012}. It produces four main X-ray emission line complexes: Cl K-$\alpha$ at 2.62~keV, Cl K-$\beta$ at 2.81~keV, K K-$\alpha$ at 3.31~keV and K K-$\beta$ at 3.58~keV, plus a small fraction of 5.89~keV backscattered X-rays. These lines are outside of the science band so they do not interfere with the supernova remnant science.

\section{Baseline Performance}
\begin{figure}
\begin{center}
\includegraphics[height=5.2 cm, keepaspectratio]{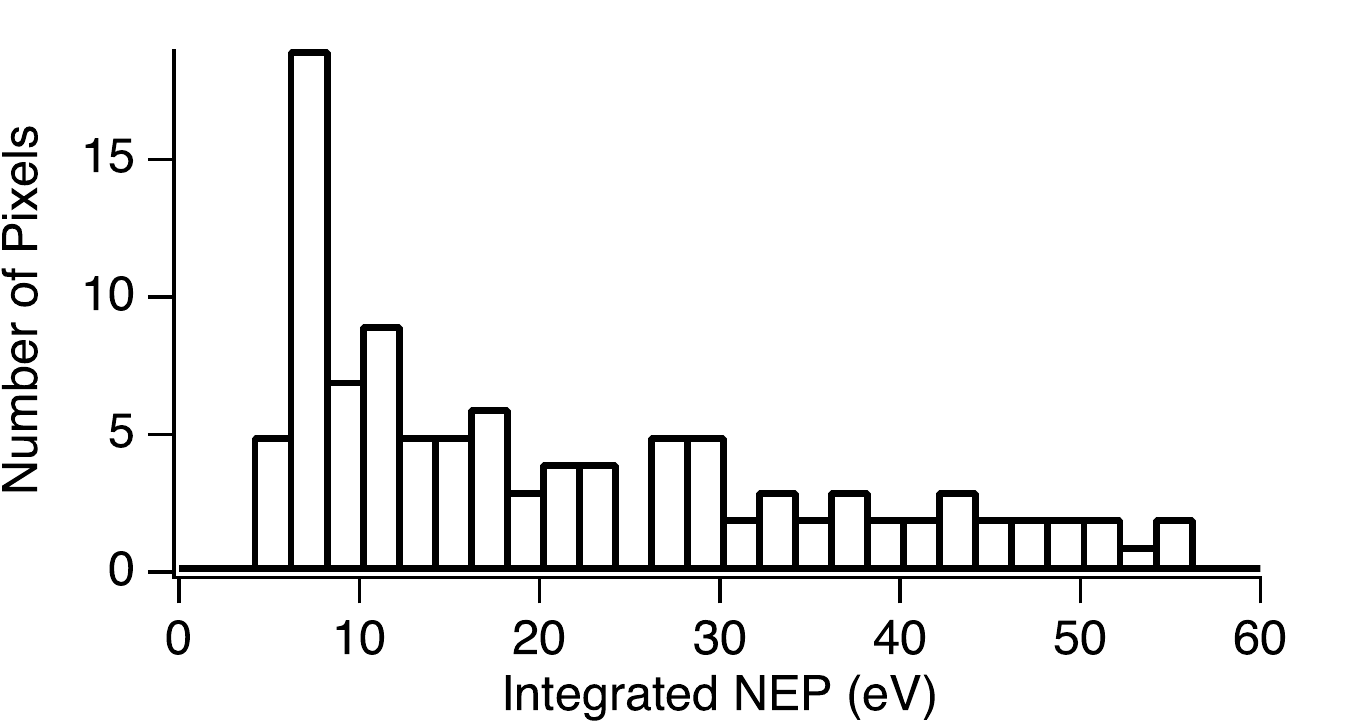}
\caption{\label{nep_hist} Histogram of the integrated NEP of different pixels within the Micro-X array while in flight configuration during preflight testing.  These resolutions are measured from the integrated NEP, which compares the detector noise spectrum to the average pulse from K K-$\alpha$ photons.  The tail of this distribution out to high NEPs is primarily due to oscillatory telemetry signal in flight configuration, for which the susceptibility varies between pixels.}
\end{center}
\end{figure}

The performance of the instrument was evaluated by measuring the integrated Noise Equivalent Power (NEP)\footnote{The integrated NEP is calculated using the optimal filter algorithm used in the XCAL GSE software discussed below, using an integral of the pulse shape and noise spectra \cite{XCalGSE}.} of all detectors/pixels in the array both prior to and after integration with the rest of the rocket payload. Out of the 128 detectors, 117 were operational. In the pre-integration tests, the instrument achieved an integrated NEP of less than 10 eV on 107 detectors with some reduction in performance due to RF pickup \cite{Goldfinger2018}. However, once fully integrated for flight, this number dropped to 39 detectors with integrated NEPs less than 10 eV. The best detector after integration had an integrated NEP of 4.4 eV (Fig. \ref{nep_hist}).

One of the major differences between the pre-integration and post-integration tests was that before integration we only operated one half of the array at a time, with flight electronics assembled on a lab rack. We call this the ``laboratory configuration''. During integration we had all the electronics in the rocket skin, and were operating the entire array at once. We call this the ``flight configuration''.  In order to maintain redundancy to failures during flight, the Micro-X detector readout is split in two halves, with each independent set of readout electronics operating on half of the array. During laboratory testing, only one side was operated at a time, while in the flight configuration, both sides were operated simultaneously.  Tests showed that the oscillators used to run digital clocks on each side of the readout, which were in close proximity to each other in the flight configuration, were beating with each other producing an oscillatory signal that impairs the performance of the array. Post-flight tests showed that this noise is removed by synchronizing the clocks to a single oscillator, using the laboratory telemetry simulator, which should result in future performance comparable to the laboratory configuration. We will be implementing this change for the next flight. 

\begin{figure}
\begin{center}
\includegraphics[height=6.2 cm, keepaspectratio]{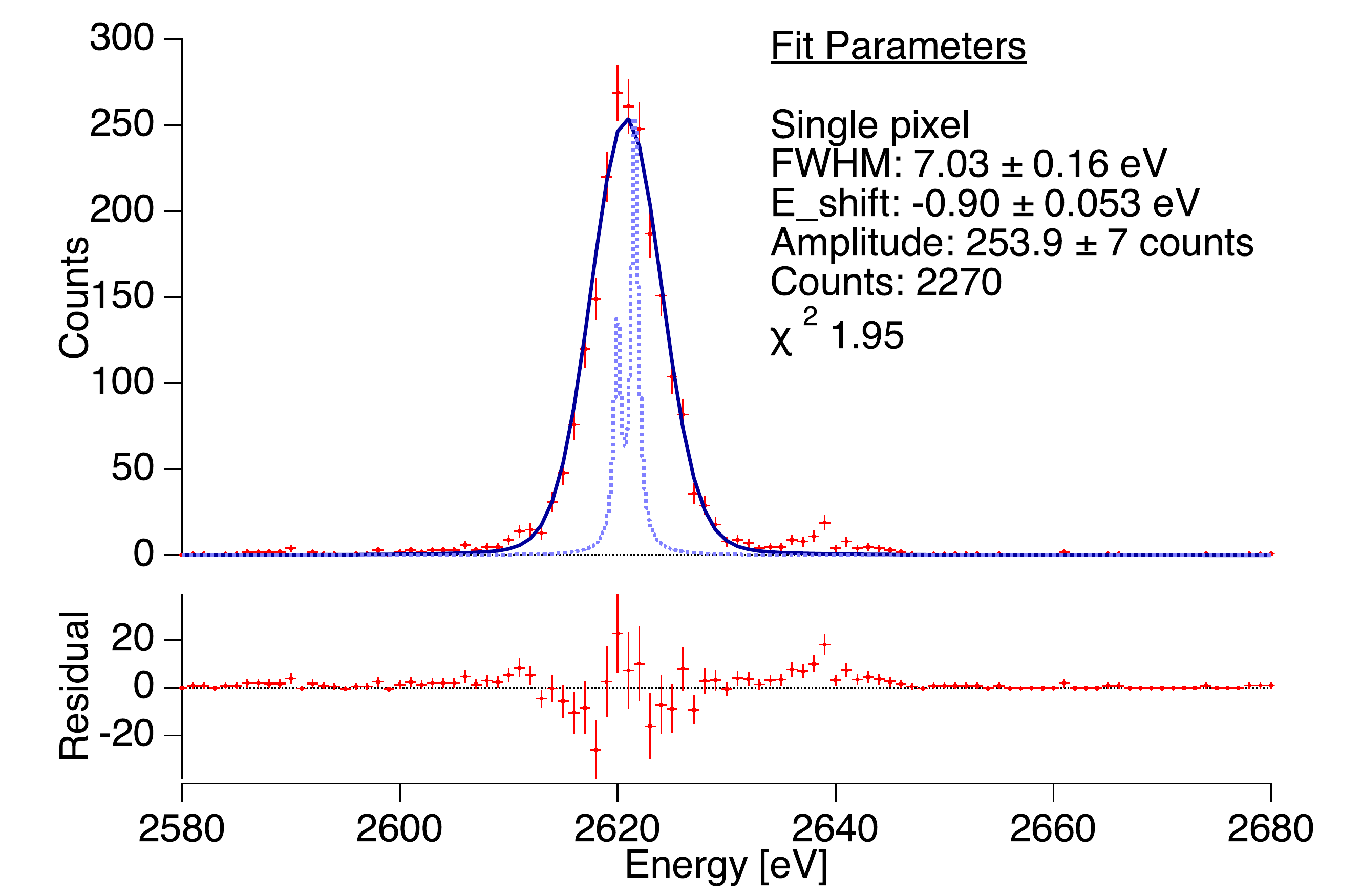}
\caption{\label{groundspec} Spectrum of the Cl K-$\alpha$ line from a single pixel in the Micro-X array, recorded during post-flight testing in the laboratory with the instrument still in the same configuration from flight.  The measured spectrum is indicated in red, with the best fit in sold blue and the inherent line shape in dashed blue.  This pixel was selected as having the best performance, so this represents the ideal energy resolution achieved by the first flight configuration of the Micro-X instrument. (Color figure online.)}
\end{center}
\end{figure}

The energy resolution of the instrument was evaluated by collecting long data sets (to collect enough statistics) and fitting the calibration lines for each detector. Past measurements with the Micro-X array have shown a difference between the integrated NEP and the energy resolution at an X-ray emission line, measured as the Full-Width at Half Maximum (FWHM) of a gaussian convolved with natural spectral shape of the emission line \cite{Goldfinger2018}. In Fig. \ref{groundspec} we show a fit to the Cl K-$\alpha$ line taken post-flight while the payload was still in flight configuration demonstrating a 7.03 eV FWHM energy resolution from a pixel with an integrated NEP of 5.97 eV FWHM.

\section{First Flight}
On the night of the July 22\textsuperscript{nd}, 2018, the Micro-X rocket payload was launched for the first time. The complete flight lasted for approximately 13 minutes. After ignition, both motors separated before the opening of the shutter door located at the aft end of the science instrument. The gate valve separating the X-ray optics from the cryostat environment was opened on a timer, at the altitude where the ambient pressure is compatible with the cryostat vacuum. This event marked the beginning of the science observation period. However, during the ascent an unexpected software reboot resulted in the loss of attitude reference and the subsequent failure of the attitude control system (ACS) to point to the desired target. This resulted in the payload slowly tumbling for the duration of the flight, with no time observing Cassiopeia A. After 320 seconds of data taking, the gate valve closed as the instrument descended into the atmosphere. A parachute located at the forward end of the payload opened after reentry, and landing occurred 780~s after ignition in the White Sands desert.

During flight, the instrument sent 42$\%$ of the science data to telemetry, which was successfully communicated to ground through an antenna system. The full 32 GB data stream was processed by the on-board electronics and was recovered after flight. All of the flight timers were successfully actuated. The rocket experienced a peak vibration load of 12.5 g during the ascent, but no temperature change was visible on the detector stage of the ADR during the launch.  The instrument launched with an ADR temperature of 300~mK. This temperature is a trade-off between increased entropy of the salt pill and therefore less susceptibility to heat loads from launch, and the time to cool down to the 75~mK operating temperature after rocket burnout for the science observation. The system achieved successful regulation 99~s after the start of temperature regulation with 7~$\upmu$K precision.  This temperature regulation error is a subdominant contribution to detector performance. However, the science observation began 51~s after the start of temperature regulation, so the temperature was still changing during the first 49~s of observation. We are optimizing our time-to-regulation parameters based on the first flight data to avoid this temperature drift for the next flight.

\begin{figure}
\begin{center}
\includegraphics[width=0.8\linewidth,keepaspectratio]{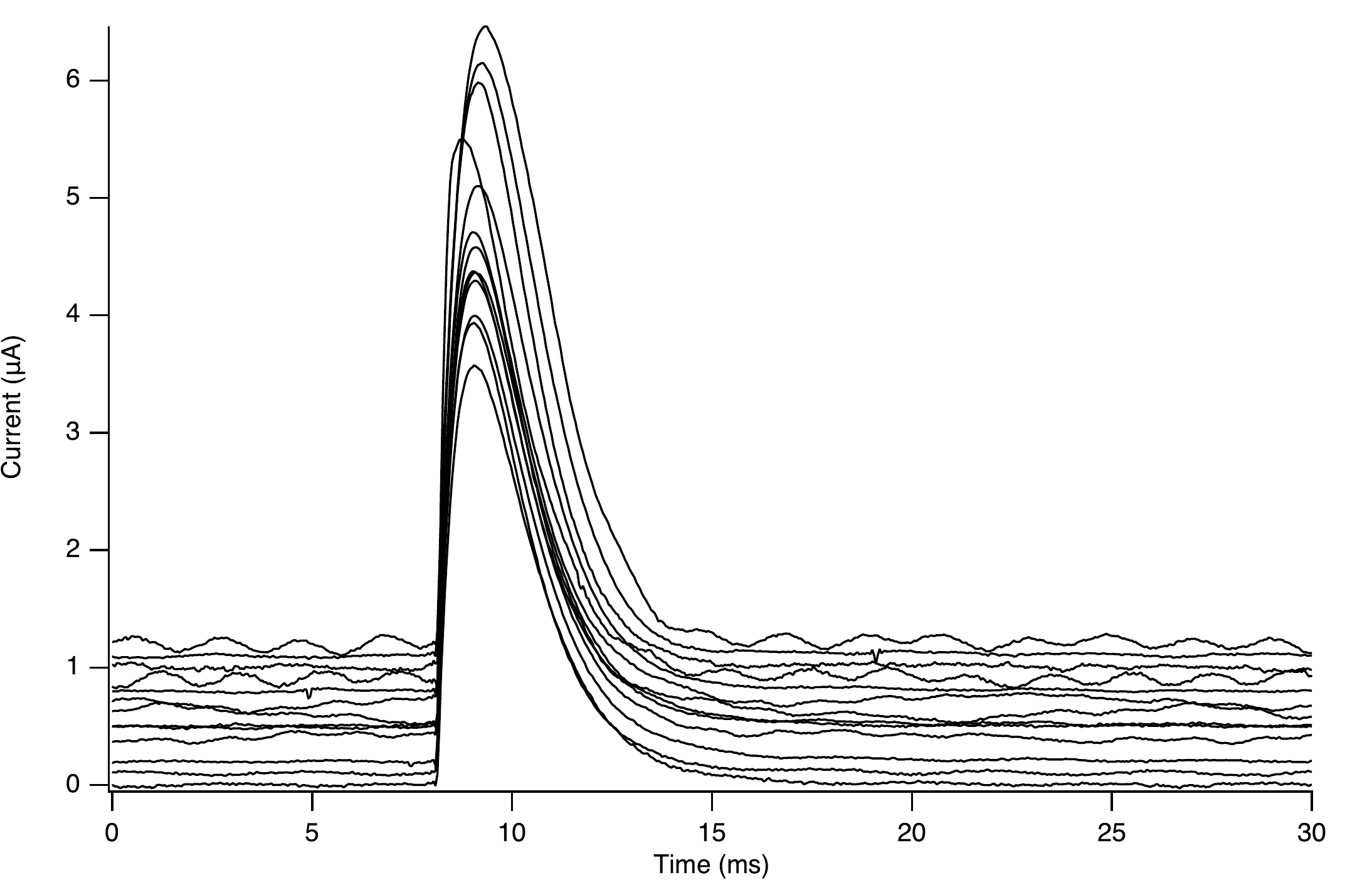}
\caption{\label{flightpulses} X-ray pulses from the K-K$\alpha$ line measured during the science observation from different pixels in a multiplexing set.  An artificial offset is added for clarity, sorted by pulse height.  The oscillatory noise can be seen in the baseline of these pulses, where the level of pickup differs between detectors.}
\end{center}
\end{figure}

\begin{figure}
\begin{center}
\includegraphics[width=0.8\linewidth,keepaspectratio]{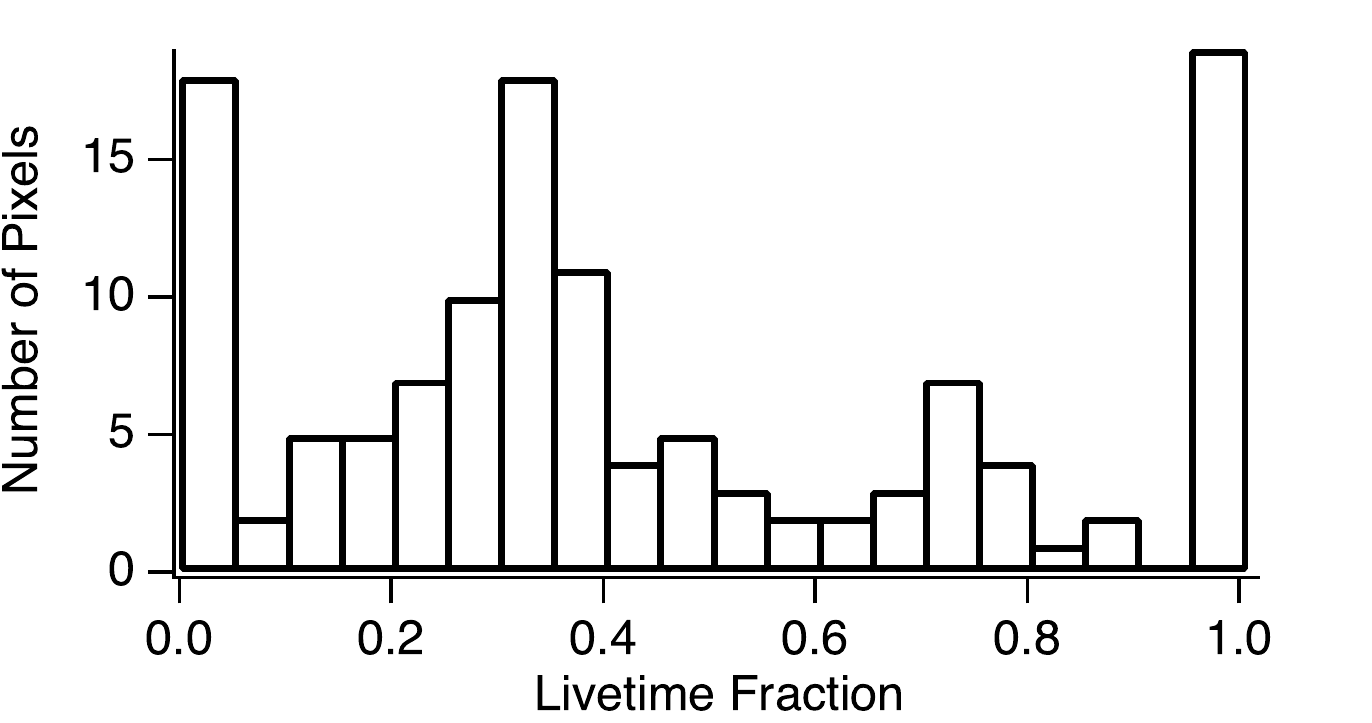}
\caption{\label{livetime} Histogram of the fractional livetime of different pixels within the Micro-X array during the portion of the flight with stable temperature used for analysis.  For the inactive pixels, the SQUID readout had lost lock due to a shift in the V/$\Phi$ response curve of the higher stage SQUIDs that is not actively corrected for changes in input.  The state of the detector is determined via the stability of the detector baseline using the algorithm discussed in Sec. \ref{sec:flight_perf}.}
\end{center}
\end{figure}

As expected due to the pointing failure, all X-rays observed during flight are consistent with coming from the on board calibration source with an average rate of $\sim0.7$ counts/pixel/s (Fig. \ref{flightpulses}).  This yields a K K-$\alpha$ rate of 0.23 +/- 0.06 counts/pixel/s. \com{Tali made this change:} No astronomical X-ray sources strong enough to accumulate a detectable number of counts passed through the FOV during the flight, so no counts from external sources were expected.

During the flight, many of the SQUID amplifiers lost the lock used to track the input signal, so their detectors could not be read out.  The amount of time that the SQUIDs were locked varied across the array (Fig. \ref{livetime}) but was similar among detectors that share the higher stage SQUIDs of their SQUID readout.  One possible explanation is the pick-up of ambient magnetic fields by the SQUID amplifiers.  The SQUIDs used in this flight are an older (2006) design that have poorer gradiometry than current designs and are sensitive to external magnetic fields \cite{Stiehl2011}.  Since the pointing failure led to tumbling, the ambient field direction from the magnetosphere with respect to the detector pointing axis changed during the flight, which is hypothesized to explain the loss and regain of lock.  This motion is not part of a nominal flight, so the magnetic shielding was designed under the assumption that the only time-varying fields would arise from the ADR magnet.  The openings in the niobium enclosure reduce the shielding factor for certain field vectors more than others, so the field would have aligned with the axis of poorer shielding as the rocket tumbled. The susceptibility is very hysteretic so while it is possible to reproduce the loss of lock with magnetic fields of similar amplitude in the lab, it cannot be precisely reproduced. The most significant effect is seen along the openings for the readout cables, so this might be improved through the addition of a high-permeability material at an outer stage of the cryostat for future flights without requiring that the cables be changed to accommodate this.

Another possibility notes that the detectors and SQUID readout are known to be susceptible to pickup of RF noise \cite{Goldfinger2018}.  Thus the rocket telemetry may have caused a problem once the instrument gate valve was opened to space during flight, as the cryostat is not a sealed RF cavity with the gate valve open.  The end result is that only a subset of detectors had a significant exposure during the observation portion of the flight, which limits the detectors available for analysis. For the second flight, a two-stage fully gradiometric SQUID system \cite{Reintsema2019} which is much less susceptible to magnetic pickup has been installed.

\section{Flight Performance}
\label{sec:flight_perf}
To generate a spectrum from the flight data we used the XCAL GSE software written for the Igor Pro environment \cite{XCalGSE}.  This allows the user to apply cuts to the data and apply an optimal filter.  It then calculates a gain calibration for the overall response, as well as a time-varying gain correction for instrument drift to generate a spectrum from the dataset.  Since there were no pulses from the astronomical source, the time stream is sparse enough that the pulses could be triggered individually with 39~ms window traces, with a small number of double pulses, defined as two pulses within the same record.  Had the target been visible, the count rate would have been high enough that an analysis of the full time stream would have been necessary to avoid pile-up.

To eliminate traces when the SQUID readout was unlocked, a cut was implemented based on the derivative of the baseline.  Any trace in which the derivative was precisely zero was unlocked because the output had railed and no data was being digitized.  Any trace with a baseline derivative greater than 1.6 A/s was also marked as unlocked.  The rapid change in output indicates that the controller was unable to establish a lock on the SQUID feedback.

For the analysis of the in-flight performance, we considered the time after the system achieved tight temperature control until the start of reentry for a net time of 317 seconds.  This excludes the first 39 seconds of the science observation when the detectors had yet to reach temperature stability and adds 36 additional seconds following the science observation when the cryogenic subsystems were still stable.  Without any time on target, the detector observed the same source after the gate valve had closed as was observed with the gate valve open.

In flight, there was a significant drift in the detector gain, to a level exceeding $2\%$ of the measured pulse height within the set of pixels with continuous readout.  This is significantly larger than the intrinsic energy resolution, so the quality of the drift correction has a significant impact on the best measurable energy resolution.  We have fit the gain's changes with time using a 4\textsuperscript{th}-order polynomial function that is unique for each pixel since there was no correlation between the time evolution of the gain of nearby pixels.  One challenge that was faced in this fit was the paucity of data, due to the short sounding rocket exposure and the low brightness of the internal calibration source during flight.  The brightness of the source was originally chosen to balance the desire for a high count rate for calibration with the need to minimize pile-up with the science data. The 2.75 half-lives elapsed since the acquisition of the calibration source led to a lower rate than would be ideal to do this calculation. A new calibration source has been acquired and will be used for the second flight.

\begin{figure}
\begin{center}
\includegraphics[height=6.2 cm, keepaspectratio]{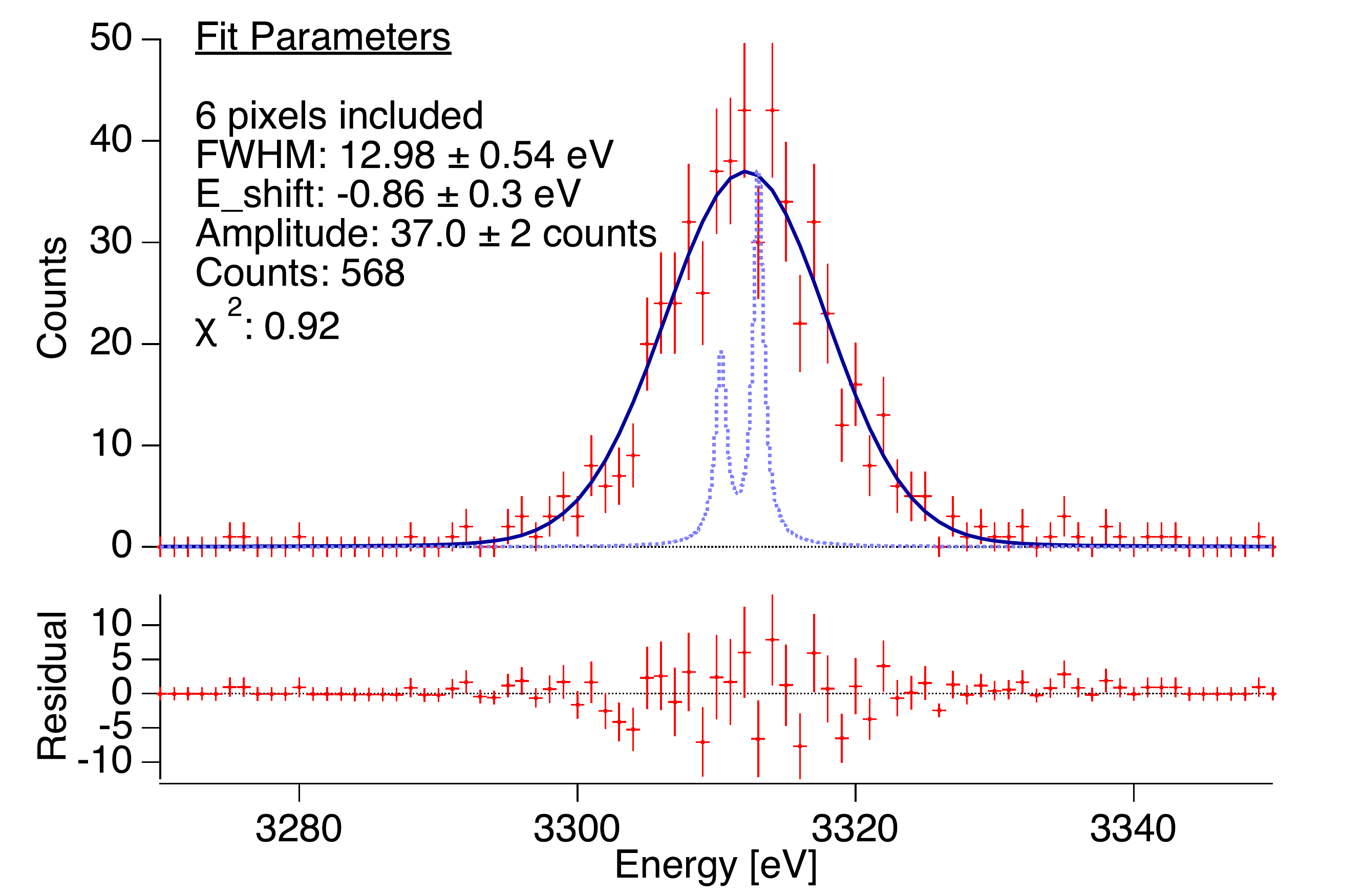}
\caption{\label{flightspec} Coadded spectrum of the K K-$\alpha$ line from a collection of 6 pixels in the Micro-X array, recorded during the analysis period of the flight.  The measured spectrum is indicated in red, with the best fit in sold blue and the inherent line shape in dashed blue.  The pixels were selected for having comparably good performance among those detectors whose SQUID readout was controlled throughout the flight.  The measurement only observes the internal calibration source due to the ACS failure, which limits the number of events in the spectrum. (Color figure online.)}
\end{center}
\end{figure}

Of the eight columns of pixels, each read out by a single second stage SQUID and SQUID array, only one column maintained lock on all pixels for the entirety of the science observation, providing 14 out of the 19 pixels that maintained consistent readout.  Since each of these channels had fewer than 110 events from the brighter K K-$\alpha$ line, we coadded the spectra of several pixels from this column to give a combined energy resolution for the array.  To demonstrate the capability of this instrument in flight, we selected the 6 pixels whose individual energy resolution was measured at better than 15.0~eV, leading to a total of 568 events which have a FWHM energy resolution of 13.0~eV at 3.3~keV (Fig. \ref{flightspec})\footnote{Additional pixels with partial data do not add sufficient counts to help the fit when compared to the worse drift correction arising from the shorter observation.}.  The remaining operable pixels are omitted due to their greater susceptibility to the oscillating signal generated by the simultaneous operation of both halves of the array in flight configuration, which leads to worse detector performance (Fig. \ref{nep_hist}).  The degradation in the energy resolution from the NEP projections here is larger than is seen on the ground, however this is to be expected given the magnitude of the detector gain drift in time.  As there are many competing variables that will determine the state of the TES, it is difficult to ascribe the drift to a single mechanism, although it is possible that the unstable motion of the sounding rocket due to the ACS failure played a role.  The temperature stability on the detector plane suggests that thermal drift was not the cause.  Gain drift of this magnitude was not seen in testing on the ground, including in pre-flight sequence testing and vibration tests that replicate the conditions during flight.

\section{Conclusion}
Despite the rocket pointing error, the Micro-X first flight was an engineering success. It demonstrated the first operation of TES and SQUID multiplexing in space. The unexpected tumbling environment limited the detector resolution and readout, which demonstrated 13.0~eV resolution from six selected pixels as compared with 7.0~eV measured in the same configuration on the ground. Full data recording, recovery and real time data transmission were successful and the ADR was able to maintain the 75~mK base temperature through the end of the observation with the required stability.  Micro-X will fly again in a similar payload configuration in December 2019 to survey the Puppis A supernova remnant. For that flight, the enclosure for the electronics is being upgraded to improve the RF shielding. The MUX system has been replaced by a new SQUID readout featuring a two-stage design with higher order gradiometry for lower susceptibility to external fields. In order to reduce the oscillating noise visible when operating both sides of the science chain, the clocks are planned on being synchronized by a single oscillator. A new stronger calibration source has been installed. These modifications address some of the leading challenges from the first flight so that the instrument can achieve its optimal performance in future flights.

\begin{acknowledgements}
The Micro-X project is conducted under NASA grant 80NSSC18K1445. Part of this work was performed under the auspices of the U.S. Department of Energy by Lawrence Livermore National Laboratory under Contract DE-AC52-07NA27344.
\end{acknowledgements}

\pagebreak

\end{document}